# Element-Specific Phonon Density of States of Iron in

# LaFeAsO$_{1-x}$F$_x$ and La$_{1-x}$Ca$_x$FePO

Satoshi HIGASHITANIGUCHI[1,2], Makoto SETO[1,2,3], Shinji KITAO[1,2], Yasuhiro

KOBAYASHI[1,2], Makina SAITO[1,2], Ryo MASUDA[2,3], Takaya MITSUI[2,3], Yoshitaka

YODA[2,4], Yoichi KAMIHARA[5], Masahiro HIRANO[5], and Hideo HOSONO[5,6,7]

[1]*Research Reactor Institute, Kyoto University, Kumatori-cho, Sennan-gun, Osaka 590-0494,*

*Japan*

[2]*CREST, Japan Science and Technology Agency, Honcho Kawaguchi, Saitama 332-0012,*

*Japan*

[3]*Japan Atomic Energy Agency, 1-1-1 Koto, Sayo-cho, Sayo-gun, Hyogo 679-5148, Japan*

[4]*Japan Synchrotron Radiation Research Institute, 1-1-1 Koto, Sayo-cho, Sayo-gun, Hyogo*

*679-5198, Japan*

[5]*ERATO-SORST, JST, in Frontier Research Center, Tokyo Institute of Technology, Mail*

*Box S2-13, 4259, Nagatsuta, Midori-ku, Yokohama 226-8503, Japan*

*[6]Materials and Structures Laboratory, Tokyo Institute of Technology, Mail Box R3-1, 4259, Nagatsuta, Midori-ku, Yokohama 226-8503, Japan*

*[7]Frontier Research Center, Tokyo Institute of Technology, Mail Box S2-13, 4259, Nagatsuta, Midori-ku, Yokohama 226-8503, Japan*

We have measured element-specific Fe-phonon densities of states (Fe-PDOS) of LaFeAsO$_{1-x}$F$_x$ ($x$ = 0, 0.11) and La$_{1-x}$Ca$_x$FePO ($x$ = 0.13) by using nuclear resonant inelastic scattering of synchrotron radiation. The Fe-PDOS of superconductor LaFeAsO$_{0.89}$F$_{0.11}$ ($T_c$ =26 K) and that of non-superconductor LaFeAsO have similar structures to both below $T_c$ (15 K) and above $T_c$ (298 K) and, therefore, fluorine doping does not have notable effect on the Fe-PDOS. As for the superconductor La$_{0.87}$Ca$_{0.13}$FePO ($T_c$ =5.4K), the entire structure of Fe-PDOS resembles with that of LaFeAsO$_{1-x}$F$_x$, but the energy of the highest peak is higher than that of LaFeAsO$_{1-x}$F$_x$. These peaks are attributed to vibrational modes between Fe and pnicogen (As and P) and the temperature-dependent energy shifts are observed for LaFeAsO$_{1-x}$F$_x$. Observed Fe-PDOS of LaFeAsO$_{1-x}$F$_x$ agrees well with an previously calculated Fe-PDOS spectrum with a first-principles calculation and shows the structural resemblance with an calculated Éliashberg function $\alpha^2 F(\omega)$ giving small electron-phonon coupling. Therefore, our results indicate that phonons are not the main contributors to the $T_c$ superconductivity of LaFeAsO$_{1-x}$F$_x$. From the experimental viewpoint, comparison of our observed Fe-PDOS and an experimentally obtained bosonic glue spectrum will be an important clue as to whether phonons are the main contributors to superconductivity in iron-pnictide superconductors.

76.80.+y, 63.10.+a, 63.20.-e, 71.90.+q

After the discovery of superconductivity in LaFePO$_{1-x}$F$_x$ [1], LaNiPO [2], and LaFeAsO$_{1-x}$F$_x$ [3], many extensive studies have been perfomed on iron-pnictide compounds. Furthermore, it was found the replacement of La with other rare-earth elements (Pr [4], Nd [5], Sm [6]) raises the transition temperature ($T_c$) above 50 K. In addition, oxygen-deficient (ReFeAsO$_{1-x}$, Re: rare-earth metal [7, 8]) and oxygen-free bi-layer (Ba$_{1-x}$K$_x$Fe$_2$As$_2$ [9]) superconductors were synthesized. Besides cuprate superconductors, this new family of iron-pnictide compounds provides another platform to explore high-$T_c$ superconductivity. As for the cuprate superconductors, the superconductivity emerges upon doping away from a magnetically ordered mother compounds. Experiments on neutron scattering [10, 11], Mössbauer spectroscopy [12, 13] and muon spin rotation [13] have revealed the magnetic order of the Fe moments in LaFeAsO, whereas magnetic order is suppressed and superconductivity is observed upon doping [10, 12]. This is the most striking similarity between cuprate and iron-pnictide superconductors. However, the relevance of the electron correlation effect is controversial [14-16] and multiband superconductivity mainly composed of five Fe-3d bands near the Fermi level is predicted from first-principles calculations [17–22] whereas cuprates have only one relevant band.

The pairing mechanism that leads to the high $T_c$ in iron-pnictide superconductors seems to be the most critical and controversial issue at present. First-principles calculations predict that the phonon mechanism is not the main mechanism which induces high $T_c$ superconductivity in iron-pnictides [16, 23, 24]. Furthermore, optical spectroscopy [25] also suggests that the main pairing glue is not provided by the electron-phonon interaction. On the other hand, the strong electron-phonon coupling of the Fe breathing mode of $LaFeAsO_{1-x}F_x$ has been pointed out [26], and the electron-phonon coupling is strong even in cuprate superconductors. Therefore, the experimental study of the phonon states in iron-pnictides is highly important. To confirm the electron-phonons interaction is the main pairing glue in conventional superconductors, the structural similarity between the phonon density of states and an Éliashberg function $\alpha^2 F(\omega)$ obtained by tunneling measurements was important [27]. So far, Raman, IR and neutron inelastic measurements [25, 28-34] have been performed to investigate the phonons. In particular, element-specific phonon densities of states of Fe is highly desirable because the bands crossing the Fermi edge are mainly composed of Fe-3d bands as expected and confirmed by photo-emission spectroscopy [34-37] and, therefore, Fe is the key element for iron-pnictide superconductivity. The nuclear resonant inelastic scattering of synchrotron radiation [39, 40] offers element (isotope)-specific phonon energy spectra. This method has been improved further to elucidate site-specific phonon densities of states [41].

We have performed nuclear resonant scattering measurements for $LaFeAsO_{1-x}F_x$ ($x$ = 0, 0.11) and have obtained temperature-dependent element-specific phonon densities of states. Furthermore, we have measured element-specific phonon density of states of superconducting $La_{1-x}Ca_xFePO$ ($x$ = 0.13). In this letter, we discuss the temperature-dependence of Fe-PDOS and the change of the spectra upon doping by comparing the Fe-PDOS and the Éliashberg function calculated from first-principles band calculations [24].

The LaFeAsO$_{1-x}$F$_x$ ($x = 0$, 0.11) and La$_{1-x}$Ca$_x$FePO ($x$=0.13) samples were synthesized by methods described in ref. [3] and ref. [38], respectively.  The quality of the samples was checked by X-ray diffraction and found to be almost single phases with small amounts of FeAs and LaAsO$_4$ in the LaFeAsO, and LaAs in the LaFeAsO$_{0.89}$F$_{0.11}$. Electrical resistivity measurements of the superconductors LaFeAsO$_{0.89}$F$_{0.11}$ and La$_{0.87}$Ca$_{0.13}$FePO gave $T_c$ values of 26 K and 5.4 K, respectively.  On the other hand, LaFeAsO did not undergo a superconducting transition but, at around 150 K, had a resistivity anomaly [3] and structural phase transition [10, 42] followed by magnetic ordering [12].  Samples of ~ 25 mg were mixed with BN and polyethylene powder, and they were pressed to form a pelletized disk with a diameter of 10 mm.

The nuclear resonant inelastic scattering experiments were performed at the nuclear resonant scattering beamline (BL09XU) and the JAEA beamline (BL11XU) of SPring-8. The experimental setup is almost same as that described in ref. [40].  The electron beam current of the storage ring was 100 mA at 8 GeV.  A double-crystal Si(111) pre-monochromator was used to handle the high heat load of undulator radiation, and the radiation was monochromatized to the band width of 2.5 meV (FWHM) with a nested high-resolution monochromator consisting of asymmetric Si(5 1 1) and asymmetric Si(9 7 5) channel-cut crystals.  The energy of the radiation was varied around the first nuclear resonant excitation energy of $^{57}$Fe (14.413 keV).  The intensity of the incident beam was monitored with an ionization chamber and a beam flux monitor.  We have measured element-specific phonon energy spectra of Fe in samples as a function of the incident X-ray energy by counting the number of delayed photons using a multielement Si-avalanche photodiode detector (APD).  Measurements were performed on $La_{0.87}Ca_{0.13}FePO$ at 298 K, on $LaFeAsO_{0.89}F_{0.11}$ at 15 K, 40 K and 298 K, and on LaFeAsO at 15 K and 298 K using a liquid-He-flow cryostat.

Measured nuclear resonant inelastic scattering spectra were converted to element-specific Fe-PDOS according to the method described in ref. [43] and these are shown in Fig.1.  Note that measured samples were also confirmed to be almost single phase and the impurity in each sample is less than 10% with Mössbauer spectroscopy [12, 45].  This ensured that the PDOS obtained were only due to the Fe atoms in iron-pnictide without resorting to the site-specific nuclear resonant scattering method [41].  In $LaFeAsO_{1-x}F_x$ PDOS, no drastic spectrum change was observed upon fluorine doping and with the temperature used.

In the spectrum of $La_{0.87}Ca_{0.13}FePO$, three main peaks are found as seen in Fig.1. The positions of low (16 meV) and middle (25 meV, labelled with * in Fig.1) energy peaks are near the corresponding low (12 meV) and middle (25 meV) energy peaks of $LaFeAsO_{1-x}F_x$ spectra measured at 298 K respectively, but the high energy peak-position at about 41 meV (labelled with ** in Fig.1)  is much higher than that in $LaFeAsO_{1-x}F_x$ at about 31 meV. It is known that vibrational frequency is proportional to the square of the mass ratio (= 1.56, As:74.92, P:30.97) assuming that the coupling constant is the same.  Since this value is close to the frequency ratio (= 1.28), the highest peak is thought to be due to coupling between Fe and pnicogen.  This assignment is supported by the first-principles band calculation [24].  It is noted that, in the case of PDOS spectra, the observed peak usually does not correspond to one phonon mode as observed in Raman or IR spectra.  In conventional phonon-mediated superconductorswith tha same electron densities at the Fermi level and the strength of electron-phonon interactions, higher phonon energy leads to to a higher $T_c$ .  Therefore, the obtained energy shift shows the opposite result to that expected from conventional phonon BCS theory by assuming the same other parameters.  However, the fact that the peak intensities of LaFeAsO$_{1-x}$F$_x$ are larger than those of La$_{0.87}$Ca$_{0.13}$FePO imply the coupling between Fe and pnicogen in LaFeAsO$_{1-x}$F$_x$ is stronger than that of La$_{0.87}$Ca$_{0.13}$FePO.

No significant change in the Fe-PDOS spectra upon F-doping was observed in LaFeAsO$_{1-x}$F$_x$.  Although LaFeAsO undergoes a structural phase transition at around 150 K [10, 42], the change is so small that the Fe-PDOS change is not thought to be present within experimental error.  As shown in Fig.2, the Fe-PDOS spectra of LaFeAsO$_{1-x}$F$_x$ agree very well with those calculated from the first-principles band calculation [24], which also calculated the electron-phonon coupling constant ($\lambda$) as 0.21 giving $T_c$ below 0.5 K and the Éliashberg function giving a $T_c$ value of 0.8 K.  Furthermore, another first-principles band calculation gives the electron-phonon coupling constant ($\lambda$) as 0.2 [23].  Comparison of the obtained Fe-PDOS and the calculated Éliashberg function [24] are also shown in Fig.2.  The similarity between our experimentally obtained PDOS and the calculated spectrum gives the reliability of the calculation, and implies that the phonon mechanism alone cannot explain the high $T_c$ value of the superconductivity in iron-pnictides.  From the experimental point of view, comparing our observed firm Fe-PDOS with glue spectra, which can be obtained with tunneling or optical measurements, provides the clear insight that the fundamental mechanism of the iron-pnictide superconductors is due to phonons or other mechanisms. In fact, bosonic spectra, which were obtained by optical spectroscopy [25] with some assumptions, are entirely different from the PDOS.

We have found the energy shifts of middle (labeled with * in Fig.1) and highest (labeled with ** in Fig.1) peaks in LaFeAsO$_{1-x}$F$_x$ PDOS accompanying the temperature changes. These changes are listed in Table. I, in which the peak positions were obtained from Gaussian fits to the peaks. These indicate the anharmonic effect around the Fe atoms. In fact, the energy shifts of A$_{1g}$ (As) and B$_{1g}$ (Fe) modes of K$_{0.4}$Sr$_{0.6}$Fe$_2$As$_2$ were observed in Raman spectra and were attributed to the anharmonic effect [31]. However, for the temperature-dependent IR active Eu mode [31], the Eu mode was found to be a harmonic effect [44] and the temperature-dependent study of neutron-weighted PDOS indicated a rather harmonic system for the parent compound BaFe$_2$As$_2$ [34]. Furthermore, the observed peak positions of the highest peaks in Fe-PDOS are lower than the position of the calculated PDOS [24] even at 15 K as observed in Fig.2.

Although substitution of La with Sm raises the $T_c$ value up to 54.6 K in iron-pnictides [6], it has not exactly been proved that phonon-mediated superconductors do not attain such a high $T_c$ value. The structural resemblance and the same temperature dependence of PDOS of non-superconducting LaFeAsO and superconducting LaFeAsO$_{0.89}$F$_{0.11}$ give us the impression that phonons are not the main mechanism for high $T_c$ value in iron-pnictide superconductor. However, if strong electron-phonon interactions are present between Fe and As, which cause the structural phase transition in LaFeAsO, high $T_c$ superconductivity may arise in $LaFeAsO_{0.89}F_{0.11}$ due to the suppression of the structural phase transition and antiferromagnetic ordering by fluorine doping. In fact, strong electron-phonon coupling has been reported [26]. Further experimental and theoretical studies are required to clarify this point.

In summary, we have measured the nuclear resonant inelastic scattering of synchrotron radiation by $LaFeAsO_{1-x}F_x$ ($x = 0$, 0.11) and $La_{1-x}Ca_xFePO$ ($x$=0.13), and have obtained the corresponding Fe-phonon densities of states. Observed Fe-PDOS of $LaFeAsO_{1-x}F_x$ agree well with the calculated Fe-PDOS [24]. The excellent agreement, which affords a ground for the small electron-phonon coupling constants $\lambda$ (0.2 [23], 0.21 [24]), the same temperature dependence and the similar structures of $LaFeAsO_{1-x}F_x$ ($x = 0$, 0.11) indicate that the high $T_c$ in iron-pnictide superconductor is not due to phonons alone. However, serious discrepancies of the Fe-moment between experiments [10, 12] and band calculations [46] require further investigation of the electronic states of iron-pnictides both experimentally and theoretically. Moreover, comparing the Fe-PDOS obtained and the precise glue bosonic spectra measured by other methods will offer experimental evidence to help judge whether the paring is due to phonons or not.

**Acknowledgements**

We would like to thank the Accelerator Group of SPring-8 for their efforts in obtaining pure bunches and all of the staff of SPring-8 for their support.  We thank

Professor S. Kishimoto for his help with the APD detectors.

0806.1869v1.

Figure captions

FIG. 1.  Element specific Fe-phonon densities of states of $La_{0.87}Ca_{0.13}FePO$ [(a) 298 K], $LaFeAsO$ [(b) 298 K, (c) 15 K], $LaFeAsO_{0.89}F_{0.11}$ [(d) 298 K, (e) 40 K, (f) 15 K] obtained from corresponding spectra of nuclear resonant inelastic scattering of synchrotron radiation.

FIG. 2.  Experimentally observed and calculated phonon densities of states. (a) Phonon density of states and (b) partial Fe phonon density of states of $LaFeAsO$ calculated by Boeri et al.[24], and element-specific Fe-phonon densities of states of (c) $LaFeAsO$ and (d) $LaFeAsO_{0.89}F_{0.11}$ measured at 15 K.

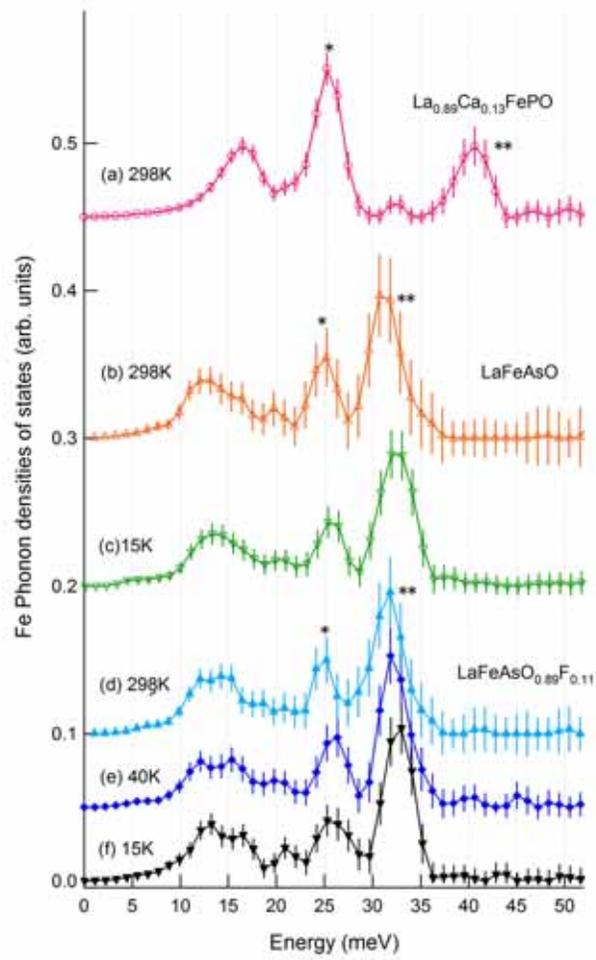

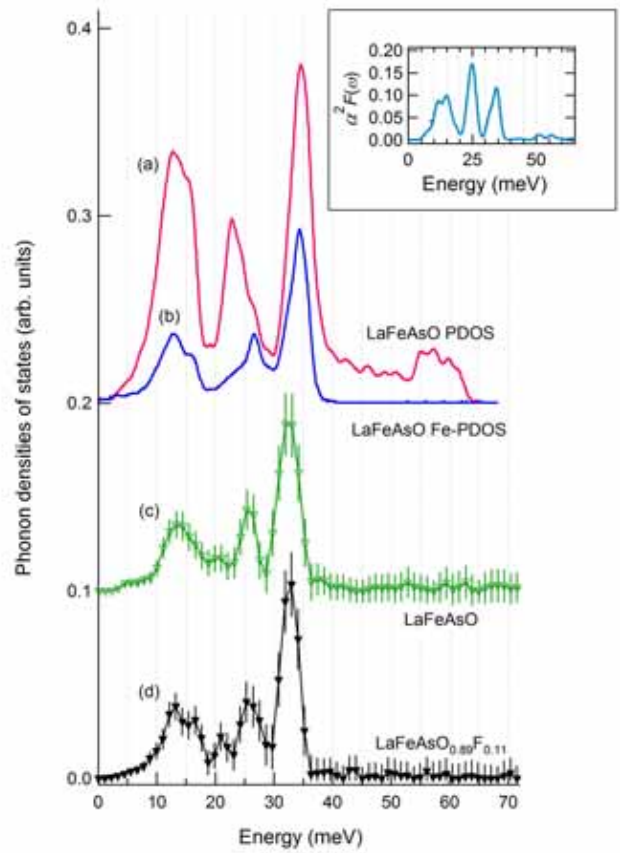

Table I. Temperature dependence of the energy shifts of peak positions observed in the PDOS spectra of La$_{0.87}$Ca$_{0.13}$FePO**,** LaFeAsO, and LaFeAsO$_{0.89}$F$_{0.11}$.  Peak 2 in this table denotes the peak position lbelled with * in Fig.1 and Peak 3 denotes the position labeled with ** in Fig.1.

| Temperature (K) | Peak positions (meV) | | |
|---|---|---|---|
| | LaFeAsO$_{0.89}$F$_{0.11}$ (Peak 2) | LaFeAsO (Peak 2) | La$_{0.87}$Ca$_{0.13}$FePO (Peak 2) |
| 298 | 24.9(4) | 24.9(5) | 25.3(1) |
| 40 | 25.9(4) | - | - |
| 15 | 25.9(5) | 25.6(4) | - |
| | (Peak 3) | (Peak 3) | (Peak 3) |
| 298 | 31.6(4) | 31.3(4) | 40.5(3) |
| 40 | 32.3(2) | - | - |
| 15 | 32.7(2) | 32.5(2) | - |